# Multiwavelength excitation Raman Scattering Analysis of bulk and 2 dimensional $MoS_2$: Vibrational properties of atomically thin $MoS_2$ layers

Marcel Placidi[1], Mirjana Dimitrievska[1], Victor Izquierdo-Roca[1], Xavier Fontané[1], Andres Castellanos-Gomez[2*], Amador Pérez-Tomás[3], Narcis Mestres[4], Moises Espindola-Rodriguez[1], Simon López-Marino[1], Markus Neuschitzer[1], Veronica Bermudez[5], Anatoliy Yaremko[6] and Alejandro Pérez-Rodríguez[1,7]

[1] Catalonia Institute for Energy Research (IREC), Jardins de les Dones de Negre 1, 08930 Sant Adrià de Besòs, Barcelona, Spain

[2] Kavli Institute of Nanoscience, T. U. Delft, Lorentzweg 1, 2628 CJ Delft, The Netherlands

[3] ICN2, Campus UAB, 08193 Bellaterra, Barcelona, Spain

[4] ICMAB-CSIC, Campus UAB, 08193 Bellaterra, Barcelona, Spain

[5] EDF R&D, 6, Quai Watier, 78401 Chatou , France

[6] Institute of Semiconductor Physics the National Academy of Science of the Ukraine, Prospect Nauki 45, Kiev 03028, Ukraine

[7] IN$^2$UB, Universitat de Barcelona, C. Martí Franquès 1, 08028 Barcelona, Spain

[*]*Present address: Instituto Madrileño de Estudios Avanzados en Nanociencia (IMDEA-Nanociencia), 28049 Madrid, Spain*

E-mail: mplacidi@irec.cat

**Abstract**

In order to deepen in the knowledge of the vibrational properties of 2-dimensional $MoS_2$ atomic layers, a complete and systematic Raman scattering analysis has been performed using both bulk single crystal $MoS_2$ samples and atomically thin $MoS_2$ layers. Raman spectra have been measured under non-resonant and resonant conditions using seven different excitation wavelengths from near-infrared (NIR) to ultraviolet (UV). These measurements have allowed to observe and identify 41 peaks, among which 22 have not been previously experimentally observed for this compound, characterizing the existence of different resonant excitation conditions for the different excitation wavelengths. This has also included the first analysis of resonant Raman spectra that are achieved using UV excitation conditions. In addition, the analysis of atomically thin $MoS_2$ layers has corroborated the higher potential of UV resonant Raman scattering measurements for the non destructive assessment of 2 dimensional $MoS_2$ samples. Analysis of the relative integral intensity of the additional first and second order peaks measured under UV resonant excitation conditions is proposed for the non destructive characterization of the thickness of the layers, complementing previous studies based on the changes of the peak frequencies.

**Keywords**: $MoS_2$, molybdenum disulfide, transition metal dichalcogenides, Raman spectroscopy, Raman resonance, multiwavelength excitation, ultraviolet Raman, atomically thin layers, few layers

## Introduction

Molybdenite or molybdenum disulfide ($MoS_2$) is one of the first two-dimensional (2D) materials which showed that graphene is not more alone competing in the world of future ultrathin 2D devices. Although mostly used as additive in oils for automotive application, $MoS_2$ is a semiconductor that has recently gained interest due to its similitude with graphene but also to its intriguing and improved properties when reduced to 2D layers. One of them, which boosted the research on $MoS_2$ and more recently enlarged on transition metal dichalcogenide (TMD) materials, has been the indirect-to-direct bandgap transition when the thickness is reduced to one single layer [1]. The presence of a direct bandgap, absent in graphene, has opened the way to a plethora of possible applications, some of them already demonstrated, such as field effect transistors [2], [3], logic operation [4], integrated circuits [5], photodetectors [6], [7], LEDs and solar cells [8], [9] using $MoS_2$ atomically thin layers.

The next promising step relies on the fabrication of 2D heterostructure devices consisting on a layer-by-layer building which recently demonstrated interesting functionalities [10]–[12].

However, constructing new 2D heterostructures directly implies to clearly identify the different materials forming the devices. Usually, the identification of layered materials is always accompanied of Raman peaks study, which also allows determining the number of layers [13]–[15]. For $MoS_2$, this has been mostly done with the standard green excitation (514-532 nm) or in resonant conditions (633 nm) [16]–[20] and the studies have been mainly centered in the analysis of the two main $E_{2g}^1$ and $A_{1g}$ Raman peaks. However, up to know there is very scarce information in the literature on the Raman scattering analysis of $MoS_2$ atomically thin layers using other excitation wavelengths [20], [21] and only some old works on bulk or $MoS_2$ nanoparticles materials can be found [22]–[27].

Li et al. [20] reported a study involving other laser lines (325 and 488 nm) for $MoS_2$ atomically thin layers with the detection of other peaks at resonance conditions. Similarly more recently a work reported on several excitation wavelengths study [21] with the detection of some peaks but unfortunately no complete experimental identification of the $MoS_2$ peaks has been performed.

The use of several excitation wavelengths allows the coupling of the Raman scattering process with real energy bands and thus also producing resonant conditions. The applicability of different (non-resonant, resonant) excitation conditions presents a strong interest from the point of view of a complete characterization of a material, allowing the enhancement of weak peaks by breaking the Raman selection rules and activating forbidden and/or inactive modes that are non-detected in standard non-resonant conditions [28].

In this framework, and to analyze the potential of Raman scattering for the assessment of atomically thin $MoS_2$ layers, a complete and systematic Raman scattering analysis has been performed using seven excitation wavelengths covering a wide spectral range from UV to IR regions. In a first stage, and in order to deepen in the knowledge of the vibrational properties of these materials, the study has been centered on bulk single crystal $MoS_2$ samples, studying the existence of a resonant excitation of several peaks when using 633 nm, 785 nm and 325 nm excitation wavelengths that are related to electron-phonon coupling with different transitions in the $MoS_2$ energy band structure. The simultaneous fitting of the spectra measured with the different excitation wavelengths has allowed completing the identification of the fundamental and second order $MoS_2$ Raman peaks, reporting and identifying 41 peaks, among which 22 have not been previously experimentally observed for this compound and characterizing the existence of different resonant excitation conditions for the different excitation wavelengths. This has also included the first analysis of resonant Raman spectra that are achieved using UV excitation conditions. For atomically thin $MoS_2$ layers, the analysis performed shows the higher potential of UV resonant Raman scattering measurements for the non destructive characterization of the samples. This is based on the strong dependence of the relative integral intensity of the additional first and second order peaks in the UV resonant Raman spectra with the number of layers.

## Results and Discussion

### Bulk $MoS_2$

Bulk 2H-$MoS_2$ belongs to the space group $D_{6h}^{4}(P6_3/mmc)$ with the unit cell containing two Mo and four S atoms. The correlation method gives the following irreducible representations of the Brillouin zone-center phonons (Γ point):

$$\Gamma \equiv A_{1g} \oplus 2A_{2u} \oplus 2B_{2g} \oplus B_{1u} \oplus E_{1g} \oplus 2E_{1u} \oplus 2E_{2g} \oplus E_{2u}$$

where $A_{1g}$, $E_{1g}$ and two $E_{2g}$ are the Raman active modes, $A_{2u}$ and $E_{1u}$ are infrared active acoustic modes; and $B_{2g}$, $B_{1u}$ and $E_{2u}$ are inactive phonon modes.

In order to understand the potential appearance of resonant excitation conditions when using different excitation wavelengths, room temperature photoluminescence measurements have been performed on the bulk $MoS_2$ single crystal. Figure 1 shows the room temperature photoluminescence measured under the 532 nm excitation wavelength. Two weak contributions are obtained at 1.84 and 2.00 eV together with a stronger one at 1.3 eV. The 1.84 and 2.00 eV emissions have been associated to the splitting of the A and B excitonic bands from the optical band gap [29], [30]. These emissions have been previously reported mainly from $MoS_2$ single layers photoluminescence experiments. Moreover, the emission of 1.3 eV is assigned to an indirect band gap transition [30]. The energy position of the different used excitation wavelengths for the Raman characterization is also shown in the upper part of Figure 1. According to these measurements, resonant excitation conditions can be approached when using the excitation wavelengths of 633 nm (relatively close to the 1.84 eV emission related to the direct bandgap) and 785 nm and 1064 nm (relatively close to the 1.3 eV emission assigned to an indirect band transition). In this last case, a decrease in the intensity is expected when using the 1064 nm excitation wavelength, because of the use of an excitation energy lower than the indirect band transition.

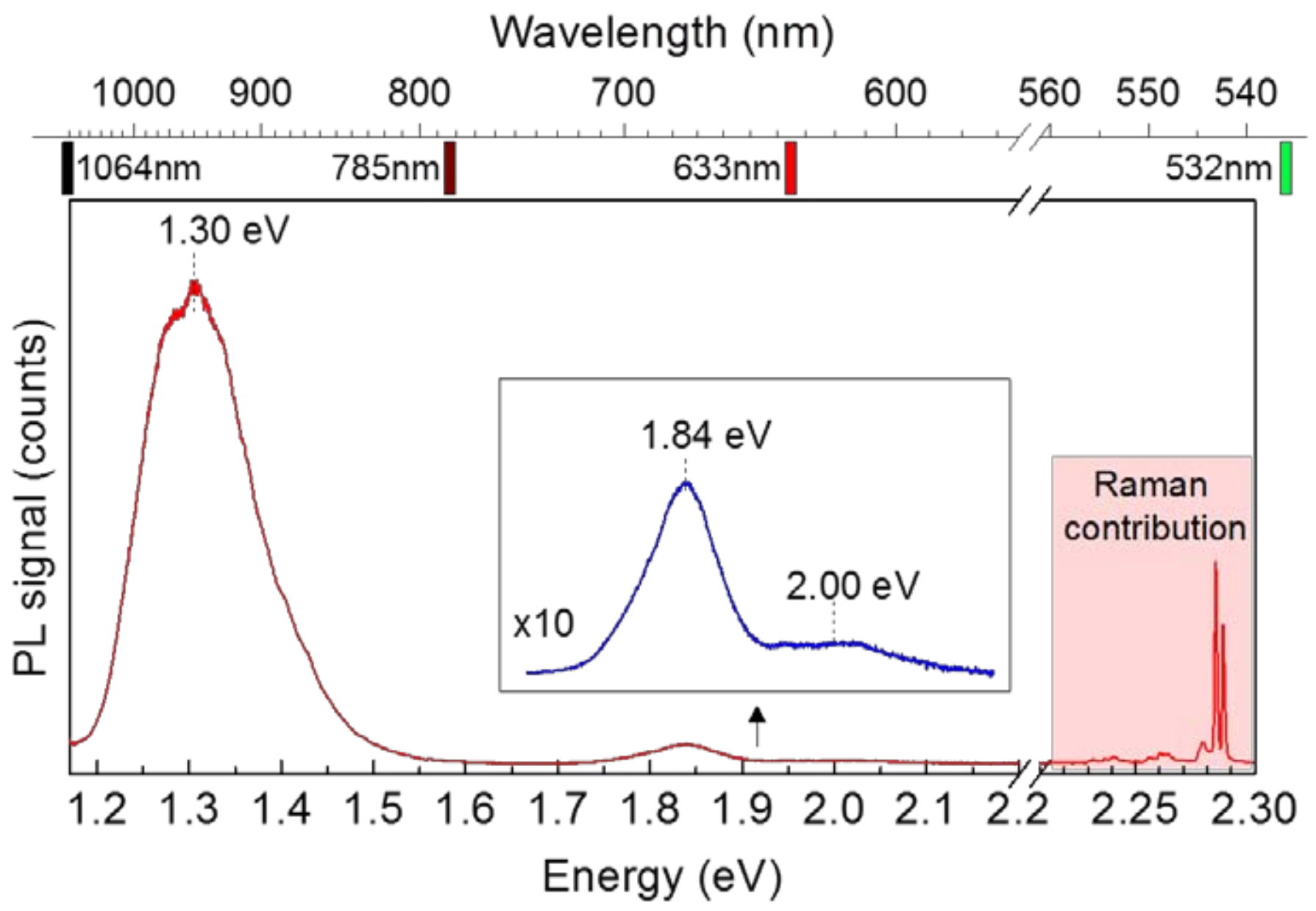

**Figure 1.** Room temperature photoluminescence spectrum of the bulk $MoS_2$ showing the two A and B excitonic peaks at 1.84 and 2.00 eV respectively, and the indirect bandgap transition at 1.3 eV. The upper part shows the energy position of the different excitation wavelengths used for Raman characterization.

Additionally, Raman measurements (which are going to be discussed later), made with 325 nm excitation wavelength also show an enhancement in the intensity of certain peaks. Resonance effects in this case are achieved because of the coupling between the excitation energy and the electronic states with energy close to 3.4 eV. Similar behavior of resonance Raman effects occurring with excitation energies different than the band-gap have also been observed in the case of kesterite materials [28]. In this case the resonance effects are due to the coupling of excitation energy with the $\Gamma_2$ point.

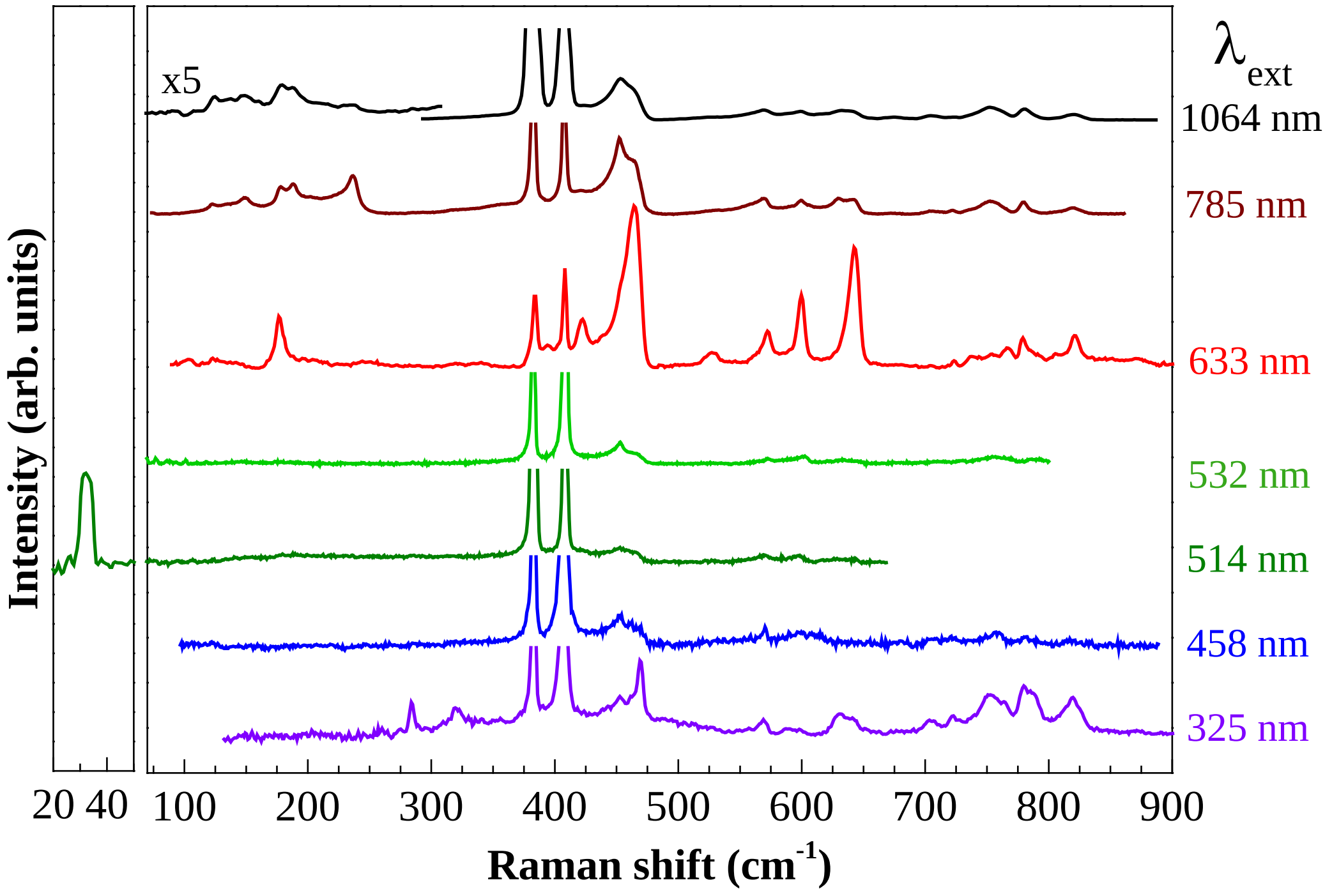

Figure 2. Raman spectra of bulk $MoS_2$ measured using different wavelength excitation, from ultraviolet (325 nm) to infrared (1064 nm).

Figure 2 shows the Raman spectra obtained under the different excitation wavelengths (from 325 to 1064 nm) using a backscattering configuration in the c-axis orientated $MoS_2$ bulk crystal. Simultaneous deconvolution of all Raman spectra with Lorentzian curves has allowed identification of 41 peaks, among which 22 have not been previously experimentally observed for this compound, and 3 peaks have been reported only for atomically thin $MoS_2$. The position of each Raman peak, the excitation condition under which it is most intense, and its proposed symmetry assignment are presented in Table I. In this table, data from the experimental peaks that are identified for the first time in single crystal $MoS_2$ are indicated in red. Table I also contains theoretical calculations [31] and previously reported experimental data [23], [22], [27] for this compound. The deconvoluted spectra for the ultraviolet (325 nm) and near-infrared (785 nm) are shown in the supporting information.

All the experimental spectra are dominated with two very intense peaks at 382.9 and 408.1 cm-1, which are attributed to $E_{2g}^1(\Gamma)$ and $A_{1g}(\Gamma)$ vibration modes [23]. Besides these, a low frequency fundamental peak is observed at 32.0 $cm^{-1}$ and identified as $E_{2g}^2(\Gamma)$ [23]. Additionally, very intense and narrow peaks are observed at 285.0 and 469.8 $cm^{-1}$ under the 325 nm excitation wavelength and are assigned to $E_{1g}(\Gamma)$ and $B_{2g}^1(\Gamma)$ symmetry modes [23]. Interestingly both modes are not expected to be observed in the standard measurement conditions. In the first case, the $E_{1g}(\Gamma)$ is forbidden in the used backscattering configuration. Observation of this peak is related to the breakdown of selection rules induced by the resonant excitation of the mode. On the other side, the $B_{2g}^1(\Gamma)$ mode is Raman inactive due to out of plane vibrations of Mo and S atoms which do not cause the change in the polarizability. Because the 325 nm excitation wavelength has a very shallow penetration depth in $MoS_2$ (around 15 nm), activation of this mode could be explained by the modification of the crystalline point-group symmetry due to a slight rearrangement of lattice structure close to the surface of the crystal. This interpretation is supported by the fact that this mode is not observed under any other resonant excitation wavelength with higher penetration depth as those achieved with the 633 nm, 785 nm and 1064 nm excitation lines. Similar effect of activation of inactive Raman modes was observed for other layered materials, as graphite [32].

In addition, weak and slightly broader contributions detected at 395.8 and 462.3 $cm^{-1}$ under 633 nm excitation could be attributed to contributions from infra-red active modes $E_{1u}(\Gamma)$ and $A_{2u}(\Gamma)$ as was explained in [22]. Activation of these modes is again explained by the use of resonant excitation conditions. Additionally, it should be mentioned another possible identification of the peak at 462.3 $cm^{-1}$, which is identified as $E_{1g}(\Gamma)$+XA band in the reference [27]. All other peaks detected in the Raman spectra measured with 633 nm excitation wavelength are assigned to second order modes as presented in Table 1. It should be noted that in all these cases the symmetry identification proposed for the second order modes is consistent with the crystal momentum conversation principle.

Furthermore it is interesting to observe several weak peaks at 149.0, 188.1, 231.9 and 237.2 $cm^{-1}$ in the spectra measured with 785 and 1064 nm excitation wavelengths (corresponding to indirect band gap resonant excitation conditions). These peaks are completely absent in the spectra measured with the other excitation wavelengths. The symmetry selection rules [23] exclude that these peaks could be assigned to one- or two-phonon allowed modes. Additionally their half-widths (8 $cm^{-1}$) are comparable with that of first-order allowed modes (6 – 8 $cm^{-1}$), suggesting that they are due to one-phonon forbidden modes. Comparison of the frequencies of the bands with the phonon dispersion simulations [31] and inelastic neutron scattering measurements [33] leads to the assignment of the four Raman structures to one phonon forbidden TA, ZA and LA modes of the M and K points around the M and K gaps. It is not so surprising to observe these modes in resonance Raman scattering near an indirect bandgap, since it is well known that the transitions in this case are dominated by two-step processes mainly involving phonons with the same symmetry of the gap [34]. The main contribution to one-phonon forbidden resonant scattering is due to a second-order extrinsic process in which a free exciton scatters inelastically the phonon via the Fröhlich electron-phonon interaction and elastically the defect via electron-impurity interaction [35]. In this case the momentum conversation law is not preserved, since the extra momentum of the phonon is being taken by the defect [36]. Defects in this case are probably planar native defects in the form of stacking faults with the direction of growth in the c axis, which are often found in layered materials due to very low formation energy [37]. Having in mind this, one-phonon indirect resonance Raman scattering could be explained by two-step process: (1) an electron excited by photon, is taking up the momentum needed to reach the allowed indirect band gap by being elastically scattered by a stacking fault; (2) the electron is inelastically scattered by a zone-edge phonon with the symmetry of the resonant gap, before recombination. This interpretation agrees with the identification proposed in the literature for the peak at 231.9 $cm^{-1}$ (which is the only one from this structure previously reported in the literature, as shown in Table 1). It is interesting to remark that the lack of observation of these features in the spectra measured in standard off-resonant conditions and in direct band-gap resonant conditions confirm the good crystal quality of the natural single crystal sample used in this study.

Finally, Table I includes also four peaks in the 760 – 820 $cm^{-1}$ second order spectral region. Even if this is the first time these peaks are observed in single crystal $MoS_2$, similar peaks identified with $E_{1g}(\Gamma)+E^2_{2g}(\Gamma)$, 2 x $E^1_{2g}(\Gamma)$ and $A_{1g}(\Gamma)+E^1_{2g}(\Gamma)$ have been previously reported in atomically thin $MoS_2$ samples [38]. In addition, there is a peak located at 780 $cm^{-1}$, labeled as “X”peak in Table I. Identification of this peak is still not clear, as any combination of first order peaks can explain its frequency. Further analysis is required, including polarization and temperature dependence measurements.

**Atomically thin $MoS_2$**

In contrast to the bulk, in the case of atomically thin $MoS_2$ samples, and due to the loss of translation symmetry along the c-axis direction, the parity of the number of atomic layers determines the space group. An odd number of atomic layers belong to the space group $D_{3h}$ symmetry without inversion symmetry, while samples with an even number of layers belong to the $D_{3d}$ symmetry with a center of inversion. Therefore, the notation of the different involved modes changes depending on the number of layers, and the representation of the zone center phonons can be written as $\Gamma_{\text{odd}} \equiv ((3N-1)/2)(A_1' \oplus E'') \oplus ((3N+1)/2)(A_2'' \oplus E')$ for an odd number of layers (N =1,3,5…) and $\Gamma_{\text{even}} \equiv (3N/2)(A_{1g} \oplus A_{2u} \oplus E_g \oplus E_u)$ for an even number of layers (N=2,4,6…). Therefore, strictly speaking, the notation of the modes for the atomically thin $MoS_2$ samples differs from that of bulk, the Raman modes with $E_{1g}(E_{2u})$, $E_{2g}^1$, $A_{1g}$ and $B_{2g}^1$ ($A_{2u}^2$) symmetries for bulk $MoS_2$ becoming $E''$, $E'$, $A_1'$ and $A_2''$ for flakes with an odd number of atomic layers and $E_g(E_u)$, $E_g$, $A_{1g}$ and $A_{1g}(A_{2u})$ for flakes with an even number of atomic layers. However, in the following text, the 'bulk' notation is maintained for simplicity.

Figure 3 shows an optical image of the flakes exfoliated from the bulk $MoS_2$ crystal and transferred onto a 285 nm thick $SiO_2$ on a Si substrate. Very large flakes were obtained due to both the high crystalline quality of the $MoS_2$ single crystal and the used exfoliation stamp-based method [39]. Different thicknesses can be easily identified by the different color contrasts [40] and are also indicated on the image showing directly the number of atomic layers (1L for a single layer, 2L for bi-layer, etc.). The thicknesses were determined by the color contrast, by Raman measurements at green excitation and by AFM measurements (shown in the Supporting Information).

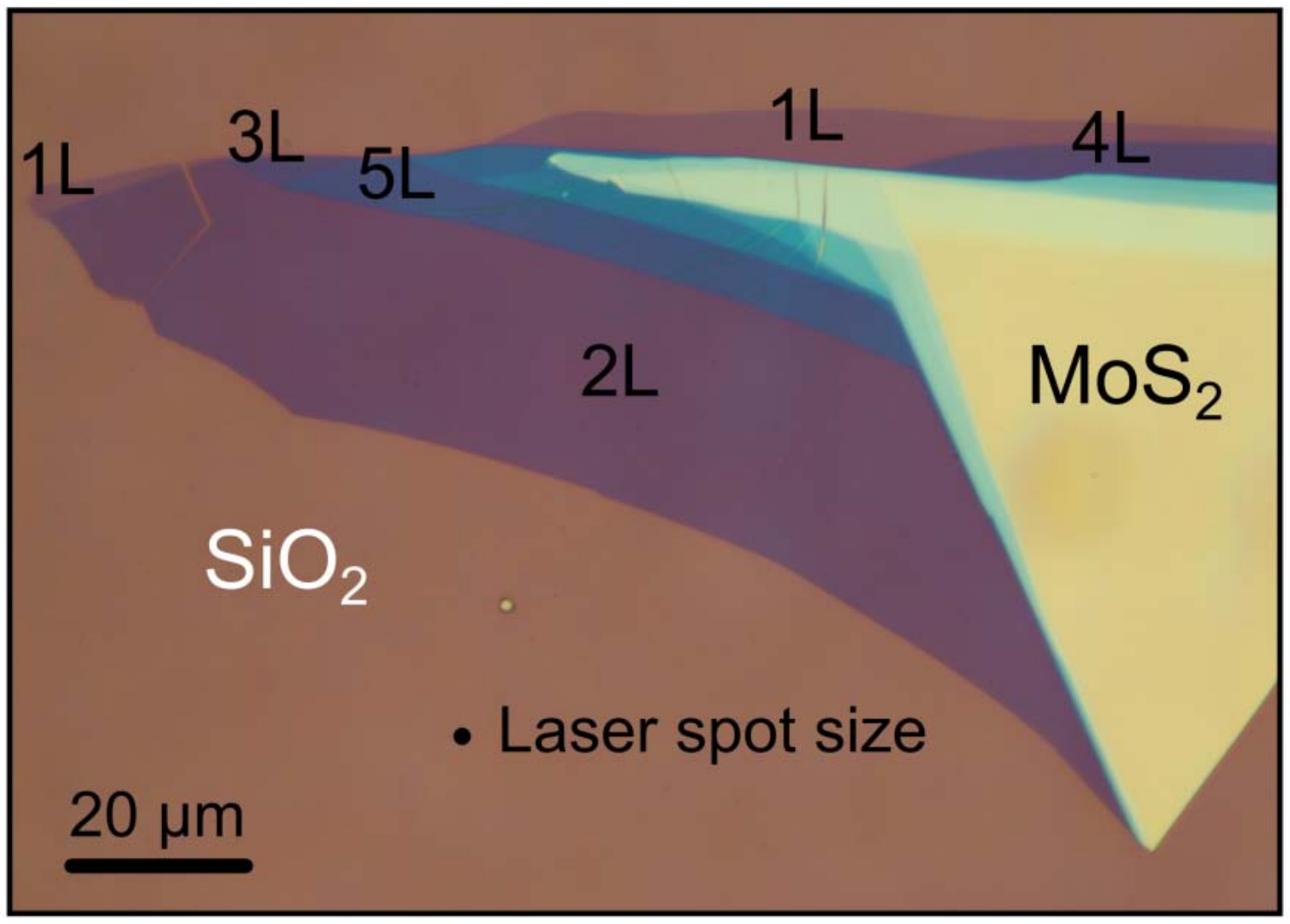

Figure 3. Optical image of a flake with single- to multilayer $MoS_2$ on 285 nm thick $SiO_2$. The number of layer (NL with N=1-6) is indicated.

Figure 4 shows the Raman spectra measured from 1L and 2L layers using UV and green excitation wavelengths. Comparison of these spectra confirms the appearance of the same additional UV resonant peaks that were observed for single crystal $MoS_2$, and identified as $E_{1g}(\Gamma)$ and $B_{2g}^{1}(\Gamma)$, as discussed in the previous section. In addition, it is also interesting to notice that the overall Raman signal obtained with the UV excitation laser line (normalized to the intensity of the Si Raman line from the Si substrate) is much stronger than when using the green excitation laser line. This indicates that Raman scattering measurements with UV excitation can be better adapted for the analysis of atomically thin $MoS_2$ layers.

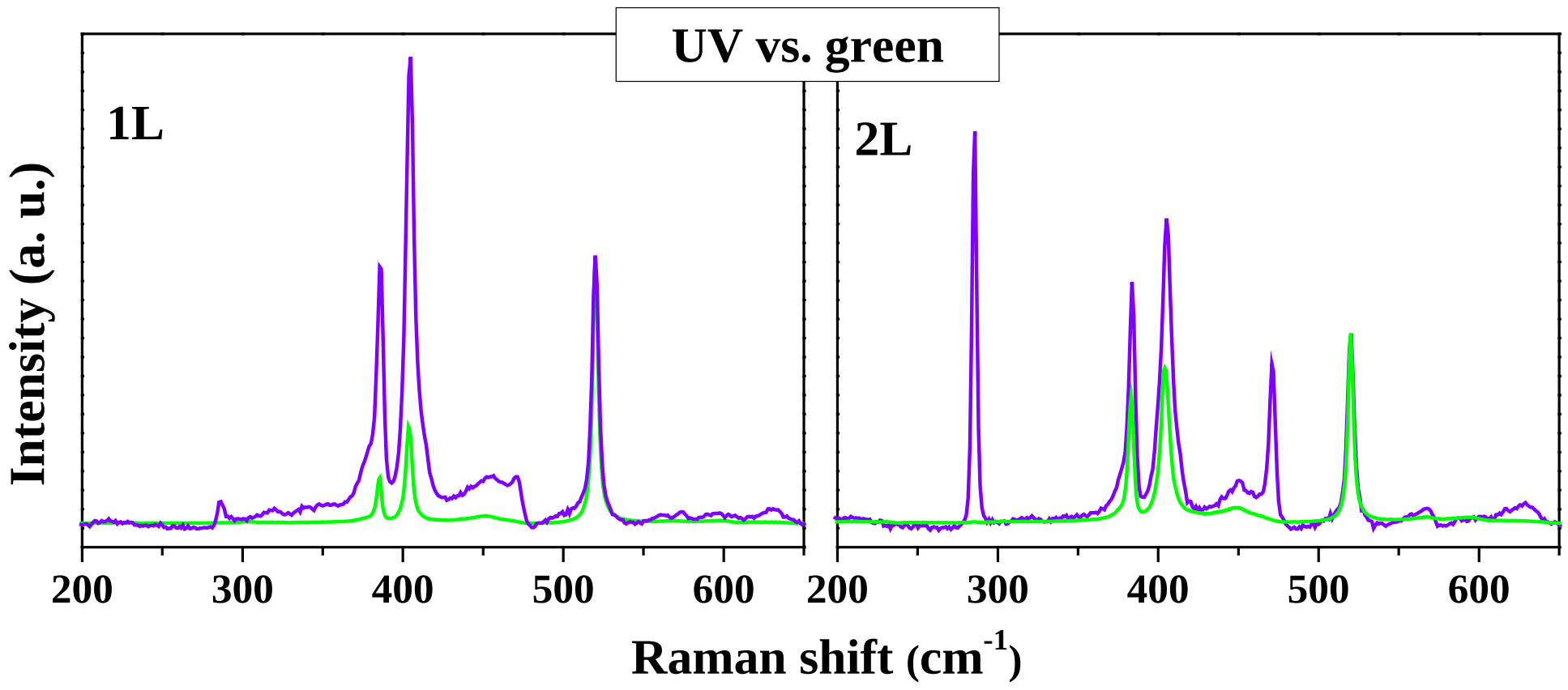

Figure 4. Comparison of Raman spectra of (a) single-layer and (b) bi-layers obtained with the green (514 nm) and ultraviolet (325 nm) excitation wavelength. The use of ultraviolet excitation allows higher signal (normalized to the Si phonon mode intensity) than when using green one.

The Raman spectra measured with UV excitation from $MoS_2$ flakes with different thicknesses are shown in Figure 5. It is interesting to note that the second order modes are particularly enhanced using the UV laser line, with intensities especially high for the single and bi-layers, as shown in Figure 5. This dramatic improvement has already been reported and was explained by the electron-two-phonon triple resonance via the deformation potential and Fröhlich interaction [38] also called triply resonant Raman scattering (TRRS).

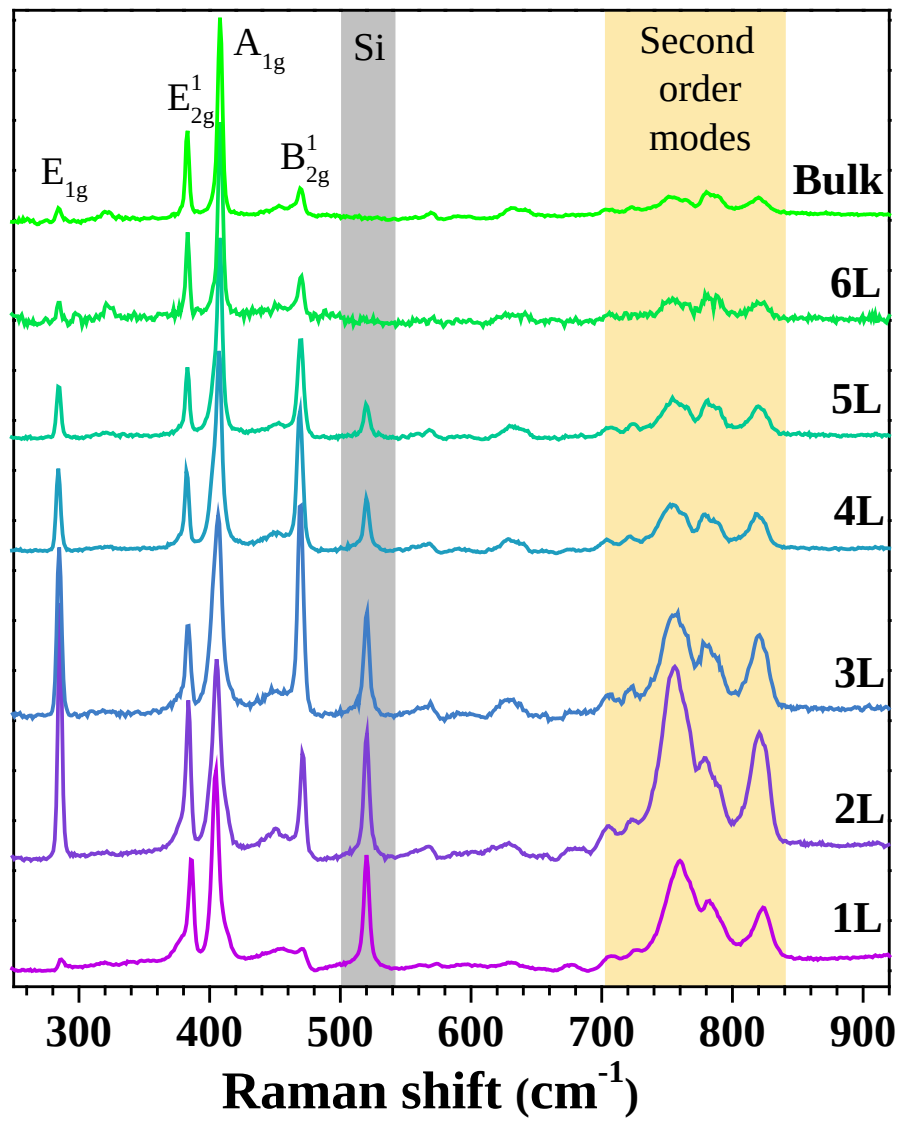

Figure 5. Raman spectra of mono- to six-layers $MoS_2$ measured with ultraviolet (325 nm) excitation wavelength. The intensities are normalized to the fundamental $A_{1g}$ mode .

Figures 6 show the dependency of the frequency of the most intense peaks obtained from the UV Raman measurements versus the number of layers. In the case of the $E^1_{2g}$ and $A_{1g}$ peaks (Fig. 6a), the observed behavior has already been reported and explained in the literature [18], [20], [29]. As shown in Fig. 6b, the two first order additional peaks appearing in the UV resonant Raman spectra, identified with the , $E_{1g}(\Gamma)$ and $B_{2g}{}^1(\Gamma)$ modes, tend to show a behavior similar to that of the $E^1_{2g}$: This is especially true for the $E_{1g}(\Gamma)$ peak, which shows a softening of ~2 cm$^{-1}$ from 1 to 4 layers and has an almost constant frequency for thicker layers, while changes in the frequency of the $B_{2g}{}^1(\Gamma)$ with the number of atomic layers are not so clear.

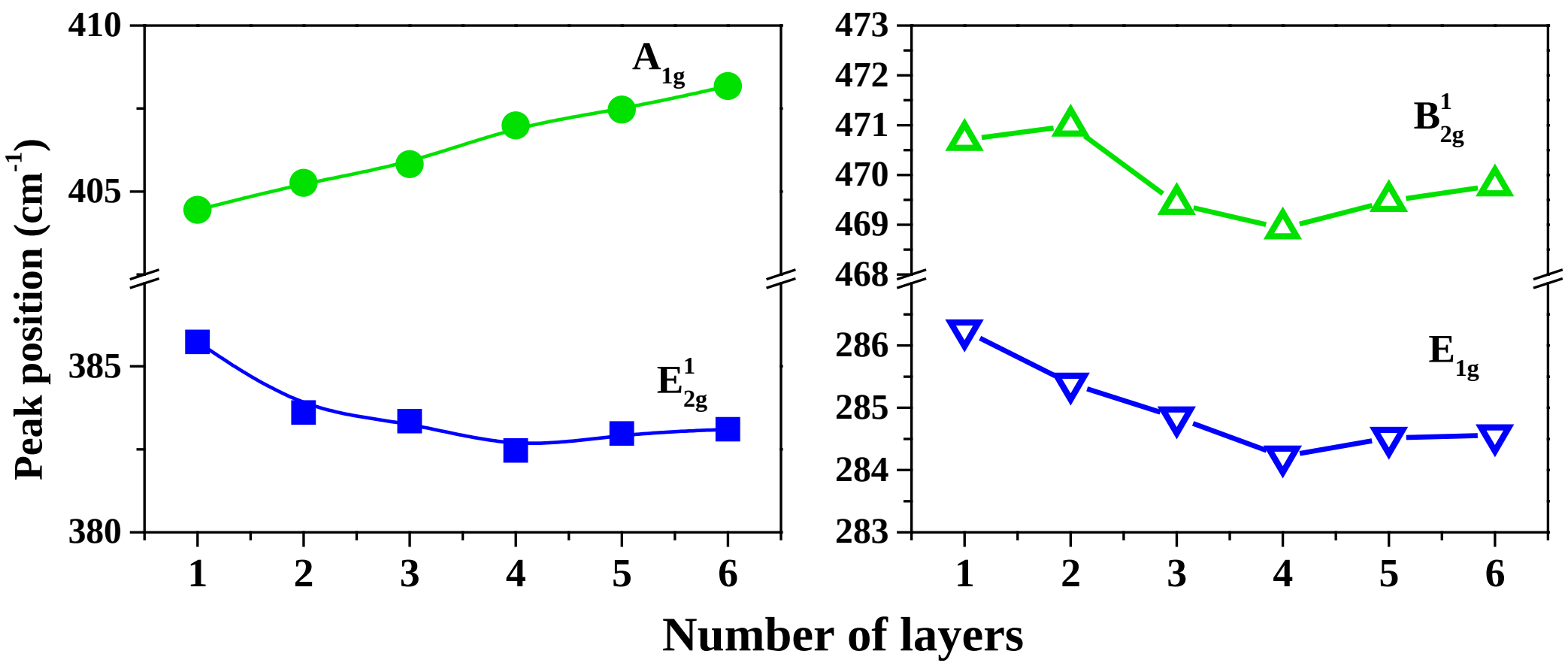

Figure 6. Peak positions of the (a) $A_{1g}$ and $E^1_{2g}$ modes, and (b) $E_{1g}$ and $B^1_{2g}$ modes as a function of the number of layers.

A much more pronounced dependence on the number of layers in the samples is observed for the relative intensity of the UV first and second order resonant peaks. Figure 7 presents the evolution of the relative integral intensity of the $E_{1g}(\Gamma)$, $B_{2g}^1(\Gamma)$ and the second order peaks in the frequency region from 700 to 860 $cm^{-1}$, calculated in relation to the main $A_{1g}(\Gamma)$ peak.

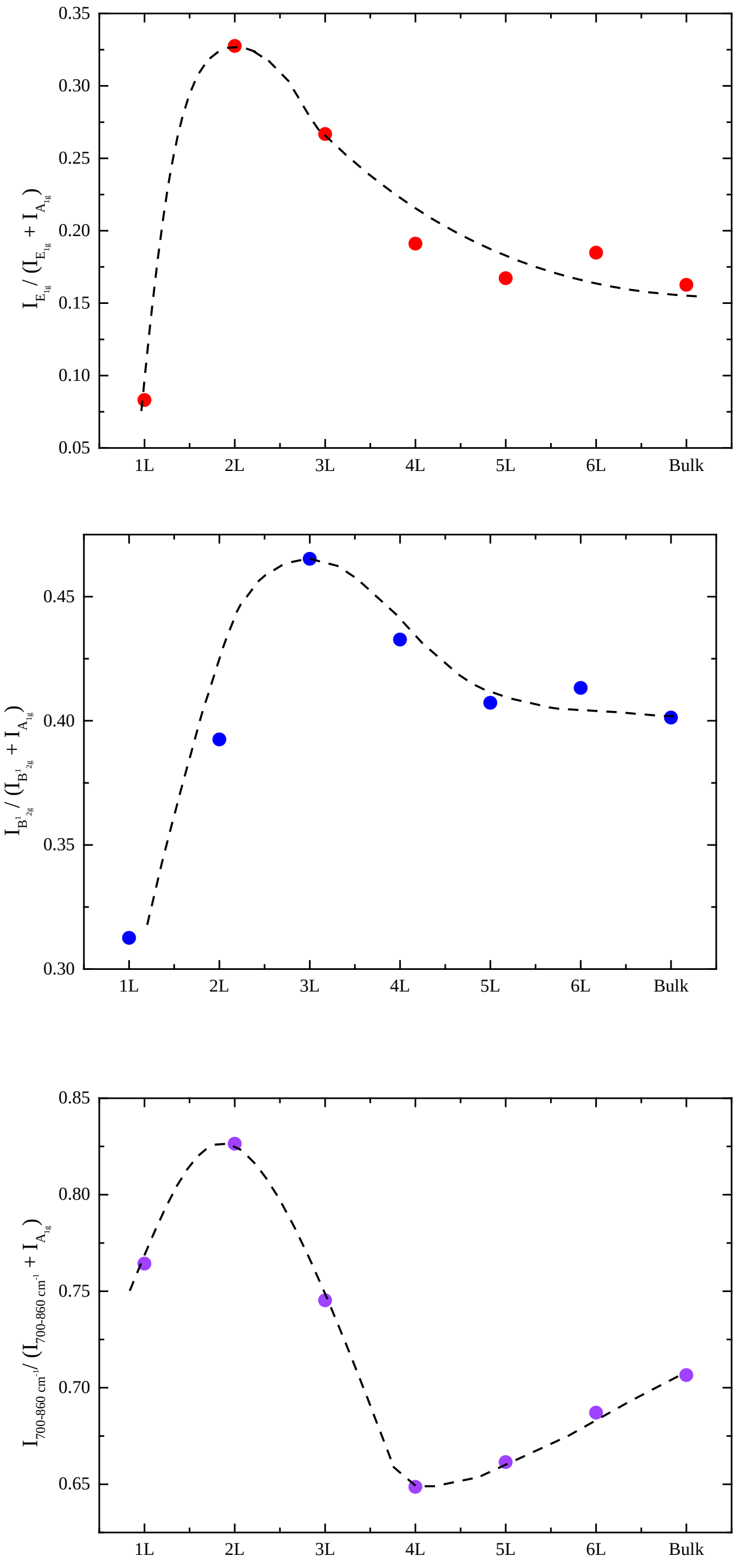

Figure 7. Evolution of the relative integral intensity of (a) $E_{1g}(\Gamma)$, (b) $B_{2g}{}^{1}(\Gamma)$ and (c) the second order peaks in the frequency region from 700 to 860 $cm^{-1}$ for the layers with different thicknesses. In all the cases, the relative intensity is calculated in relation to the main $A_{1g}(\Gamma)$ peak (corresponding to the green region indicated in the spectra shown in the insets)

From Figure 7(a) it can be observed that the relative intensity of the $E_{1g}(\Gamma)$ mode shows a drastic increase (in a factor of 4) when moving from L = 1 to L = 2. For higher values of L, the relative intensity of this mode monotonously decreases with the increase in the number of layers. A similar behavior is observed for $B_{2g}^{1}(\Gamma)$ mode, except that in this case the maximum intensity is achieved for L = 3 as presented in Figure 7(b). On the other hand, the very low intensity of these peaks in the spectra from the single layer sample contrast with the enhancement in the relative intensity of the second order peaks observed for samples with L between 1 and 3 in relation to thicker samples, as can be observed in Figure 7 (c).

The very low intensity of the $E_{1g}(\Gamma)$ mode, detected in the case of the samples with L = 1, is likely due to the interaction of the $MoS_2$ layer with the substrate. These layers are only formed by a layer of Mo atoms in between two layers of S atoms. The $E_{1g}(\Gamma)$ mode involves in-plane vibrations of only S atoms in the S layers [41]. Due to the lattice mismatch between the $MoS_2$ and the $SiO_2$ layer from the substrate, the bottom layer of S atoms, close to the substrate, has different bond lengths and angles than the surface layer of S atoms. This will essentially change the vibrations of the S atoms, which will result in a broadening and a very low intensity of the $E_{1g}(\Gamma)$ mode. This effect is only present in the case of mono layers, since already with the addition of another layer of $MoS_2$, there are already 4 S atom layers, which will lead to proper vibrations of S atoms in the case of $E_{1g}(\Gamma)$ mode, becoming these effects negligible. Similar effect is observed for the $B_{2g}^{1}(\Gamma)$ mode, except that in this case the maximum intensity is achieved for L = 3. The difference in the number of layers for which the intensity mode maximum is achieved can be explained taking into account that the $B_{2g}^{1}(\Gamma)$ mode involves out of plane vibrations of both Mo and S atoms in the lattice [41], meaning that the substrate effects are going to be more pronounced in the case of L=1 and L=2. The decrease in the intensity of these modes for L > 3 would be mainly due to the changes in the electronic structure with the layer thickness, as reported in [42]. In addition, a lowering of the intensity of these modes is also expected as the symmetry of the bulk is achieved, taking into account that they correspond to quasi-forbidden or inactive Raman modes, in agreement with their low relative intensity in the UV resonant Raman spectra from single crystal $MoS_2$ (as shown in Figure 2). On the other hand, the enhancement in the relative intensity of the second order peaks for layers with $L \leq 3$ has been interpreted in terms of TRRS effect [38], as previously indicated.

The analysis of the intensity and the frequencies of the Raman modes from atomically thin $MoS_2$ layers have been proposed in the literature for the non destructive assessment of the number of atomic layers in 2D samples, as reported in [18], [42]. In these cases, non resonant (green) or resonant (red) excitation wavelengths have been used. However, it should be mentioned that frequency of the modes are quite sensible to stress effects, which can occur due to the substrate interaction with the material, and in which case the method dependent on the frequencies shift could be corrupted. On the other side, the use of the relative intensities exclude these effects and this kind of method requires a simpler processing of the experimental data. However, the measurements performed with green (off-resonance) excitation conditions do not allow to distinguish the thickness of layers with $L \geq 2$. Use of resonant red excitation allows to achieve a higher sensitivity, being possible to detect differences between layers with L = 1, L = 2 and $L \geq 3$. The problem with this kind of analysis is fast saturation of the relative intensity with the number of layers. This contrasts with the behavior observed for the UV resonant peaks: in this case the differences in the dependence of the relative intensity of the different peaks with the number of layers allows to make a direct assessment of the thickness for layers with L = 1, L =2,

L = 3 and≥L4, increasing the range of thicknesses that can be assessed from these measurements. This analysis does not require for a fitting of the experimental spectra, being based in the analysis of the integral intensity of the experimental spectra at the different spectral regions. These data corroborate the high potential of UV resonant Raman scattering measurements for the characterization of atomically thin $MoS_2$ layers.

## Conclusions

In summary, a complete vibrational characterization of the $MoS_2$ single crystal is here reported using multi-wavelength excitation conditions. The analysis performed has allowed in a first step to complete the behavior of the vibrational properties of single crystal $MoS_2$, reporting and identifying 41 peaks, among which 22 have not been previously experimentally observed for this compound. This has been based on the simultaneous detailed fitting of the spectra measured with different excitation wavelengths covering a wide spectral range from UV to IR regions. This includes also the first report on the analysis of resonant excitation peaks that are detected using UV (325 nm) excitation conditions. For atomically thin $MoS_2$ layers, the enhancement in the relative intensity of the additional peaks measured with UV resonant excitation conditions for very low layer thicknesses gives additional interest to these measurements for the non destructive assessment of 2 dimensional layers. These results are particularly relevant and open the way to a better understanding of the vibrational properties and electronic structure in transition metal dichalcogenide materials and also to employ Raman spectroscopy to unambiguously determine the number of layers in atomically thin $MoS_2$.

## Methods

**Bulk and atomically thin $MoS_2$.** A bulk 2H-$MoS_2$ crystal (from SPI, USA) was used for all the measurements. Few layers were exfoliated from this sample and transferred onto a 285 nm thick $SiO_2$ on Si substrate using a stamp-based method[45] which allows obtaining large area flakes, as seen in the Figure 2(a). The thickness of the flakes was identified by optical contrast, atomic force microscopy and Raman measurements.

**Raman Spectroscopy.** Two different equipments were used for the Raman scattering measurements. For the 325, 532, 633 and 785 nm wavelength excitations, the measurements were performed in air in backscattering configuration with a Horiba Jobin Yvon LabRam HR800-UV system. The reduced spot size used in the microscope configuration has a diameter around 1-2 µm. This system also allows the use of a DuoScan™ accessory which allows measuring areas of approximately 30×30 µm². For the 457.9 and 514.5 nm wavelengths, the measurements were performed using a T64000 Horiba Jobin-Yvon spectrometer, in backscattering configuration, in two modes, macro mode with a spot of 100 µm on the sample, and micro mode with a spot of 1-2 µm on the sample. The Raman measurements of the bulk sample were performed in the macro mode using the T64000 system, and utilizing the DuoScan™ accessory in the case of HR800-UV system. On the other side, the Raman measurements of the few layers $MoS_2$ were performed in the micro mode which allows high spatial resolution of the order of 1-2 µm. In both cases, in order to minimize possible thermal effects, the excitation power was in the range 1-3 mW for the bulk $MoS_2$ and fixed at 0.3 mW for the layers. The first order Raman spectrum of a Si monocrystal sample was also measured as a reference sample before and after each acquisition of Raman spectra, and the $MoS_2$ spectra then corrected with respect to the Si line at 520 $cm^{-1}$.

## Acknowledgement

The research leading to these results has received funding from the People Programme (Marie Curie Actions) of the European Union's Seventh Framework Programme FP7/2007–2013/ under REA grant agreements n°269167 (PVICOKEST) and n°316488 (KESTCELLS). The research was also partially supported by MINECO, project SUNBEAM (ref. ENE2013-49136-C4-1-R ) and by European Regional Development Funds (ERDF, FEDER Programa Competitivitat de Catalunya 2007–2013). Authors from IREC and the University of Barcelona belong to the M-2E (Electronic Materials for Energy) Consolidated Research Group and the XaRMAE Network of Excellence on Materials for Energy of the "Generalitat de Catalunya". V. I.-R. thanks MINECO for the "Juan de la Cierva" fellowship (ref. JCI-2011-10782) and M.P. for the "Formación Posdoctoral" fellowship (FPDI-2013-18968).

Table I. Frequency (in $cm^{-1}$) of peaks from simultaneous fitting of Raman spectra measured with different excitation wavelengths, excitation condition under which the peak is best resolved, and proposed mode symmetry assignment. These are compared with theoretical predictions [31] and other reported experimental data [23, 22, 27].

This work

| $\lambda^a$ [nm] | $RS^b$ [$cm^{-1}$] | $Sym^c$ | $RS^d$ [$cm^{-1}$] | $RS^e$ [$cm^{-1}$] | $RS^f$ [$cm^{-1}$] | $RS^g$ [$cm^{-1}$] |
|---|---|---|---|---|---|---|
| **514.5** | 32 | $E_{2g}^2(\Gamma)$ | | | 32 | 35.2 |
| **1064** | 122.7 | $A_{1g}(\Gamma)$- $E_{1g}(\Gamma)$ | | | | |
| **1064** | 133.9 | $E_{1g}(M)$- ZA(M) | | | | |
| **1064** | 140.7 | $E_{1g}(M)$- TA(M) | | | | |
| **785** | 149.0 | TA(M) | | | | 156.0 |
| **633** | 177.9 | $A_{1g}(M)$-LA(M)* | | 177 | 178 | |
| **785** | 188.1 | ZA(M) | | | | 182.0 |
| **633** | 204.3 | $E_{1u}(M)$- TA(M) | | | | |
| **785** | 219.6 | $A_{1g}(M)$-ZA(M) | | | | |
| **785** | 231.9 | LA(M) | | 226 | 232.5 | 234.0 |
| **785** | 237.2 | LA(K) | | | | 238.0 |
| **325** | 285.0 | $E_{1g}(\Gamma)$ | 287 | 283 | 286 | 288.7 |
| **325** | 319.4 | $E_{1g}(M)$ | | | | 311.8 |
| **325** | 338.5 | $E_{2u}(M)$ | | | | 335.3 |
| **785** | 357.4 | $E_{1u}(M)$ | | | | 359.7 |
| **325** | 368.0 | $E_{2g}^1(M)$ | | | 367.5 | 372.1 |
| **633** | 382.9 | $E_{2g}^1(\Gamma)$ | 383 | 382 | 384 | 387.8 |
| **633** | 395.8 | $E_{1u}(\Gamma)$ | | | | 391.2 |
| **633** | 408.1 | $A_{1g}(\Gamma)$ | 409 | 407 | 409 | 412.0 |
| **633** | 422.3 | b' mode | | 421 | 420 | |
| **633** | 441.0 | $A_{1g}(\Gamma)$+ $E_{2g}^2(\Gamma)$ | | | | |
| **785** | 452.5 | 2 x LA(M) | | 452 | 460 | |
| **633** | 462.3 | $A_{2u}(\Gamma)$ or $E_{1g}(\Gamma)$+XA | 470 | | | 469.4 |
| **325** | 469.8 | $B_{2g}^1(\Gamma)$ | | | | 473.2 |
| **325** | 492.1 | Edge phonon | | 496 | | |
| **325** | 511.4 | $E_{1g}(M)$+TA(M) | | | | |
| **633** | 528.0 | $E_{2u}(M)$+ ZA(M) | | | | |
| **785** | 555.8 | $E_{1g}(M)$+LA(M) | | 526 | | |
| **633** | 568.7 | 2 x $E_{1g}(\Gamma)$ | | 572 | 572 | |
| **633** | 599.6 | $E_{2g}^1(M)$+LA(M) | | 599 | 600 | |
| **325** | 630.3 | 2 x $E_{1g}(M)$ | | | | |
| **633** | 641.6 | $A_{1g}^1(M)$+LA(M)* | | 641 | 643 | |
| **1064** | 675.2 | $E_{1g}(\Gamma)$+$E_{2g}^1(\Gamma)$ | | | | |
| **325** | 705.6 | $A_{1g}(\Gamma)$+ $E_{1g}(\Gamma)$ | | | | |
| **633** | 723.1 | $A_{1g}(M)$+ $E_{1g}(M)$* | | | | |
| **633** | 736.0 | 2 x $E_{2g}^1(M)$ | | | | |
| **325** | 752.0 | $E_{1g}(\Gamma)$+$E^2_{2g}(\Gamma)$ | | | | |
| **633** | 766.4 | 2 x $E^1_{2g}(\Gamma)$ | | | | |
| **633** | 780.0 | X | | | | |
| **325** | 788.6 | $A_{1g}(\Gamma)$+$E^1_{2g}(\Gamma)$ | | | | |
| **633** | 820.4 | 2 x $A_{1g}(\Gamma)$ | | | | |

$\lambda^a$ is the excitation wavelength
$RS^b$ is the Raman shift from this work
$Sym^c$ is the symmetry proposed in this work
$RS^{d,e,f,g}$ are the Raman shift reported in references [23], [22], [27], [31]
*$A_{1g}(M)$ mode has the same frequency as $A_{1g}(\Gamma)$ mode [31]

# Supporting Information:

# Multiwavelength excitation Raman Scattering Analysis of bulk and 2 dimensional $MoS_2$: Vibrational properties of atomically thin $MoS_2$ layers

Marcel Placidi[1], Mirjana Dimitrievska[1], Victor Izquierdo-Roca[1], Xavier Fontané[1], Andres Castellanos-Gomez[2*], Amador Pérez-Tomás[3], Narcis Mestres[4], Moises Espindola-Rodriguez[1], Simon López-Marino[1], Markus Neuschitzer[1], Veronica Bermudez[5], Anatoliy Yaremko[6] and Alejandro Pérez-Rodríguez[1,7]

[1] Catalonia Institute for Energy Research (IREC), Jardins de les Dones de Negre 1, 08930 Sant Adrià de Besòs, Barcelona, Spain

[2] Kavli Institute of Nanoscience, T. U. Delft, Lorentzweg 1, 2628 CJ Delft, The Netherlands

[3] ICN2, Campus UAB, 08193 Bellaterra, Barcelona, Spain

[4] ICMAB-CSIC, Campus UAB, 08193 Bellaterra, Barcelona, Spain

[5] EDF R&D, 6, Quai Watier, 78401 Chatou , France

[6] Institute of Semiconductor Physics the National Academy of Science of the Ukraine, Prospect Nauki 45, Kiev 03028, Ukraine

[7] IN$^2$UB, Universitat de Barcelona, C. Martí Franquès 1, 08028 Barcelona, Spain

[*] *Present address: Instituto Madrileño de Estudios Avanzados en Nanociencia (IMDEA-Nanociencia), 28049 Madrid, Spain*

E-mail: mplacidi@irec.cat

## Raman spectra

The deconvoluted spectra obtained for the ultraviolet (325 nm) and near-infrared (785 nm) excitation wavelengths are shown in the figure S1 and S2 respectively. Only the Raman peak positions which are the most intense under the excitation wavelength are here reported.

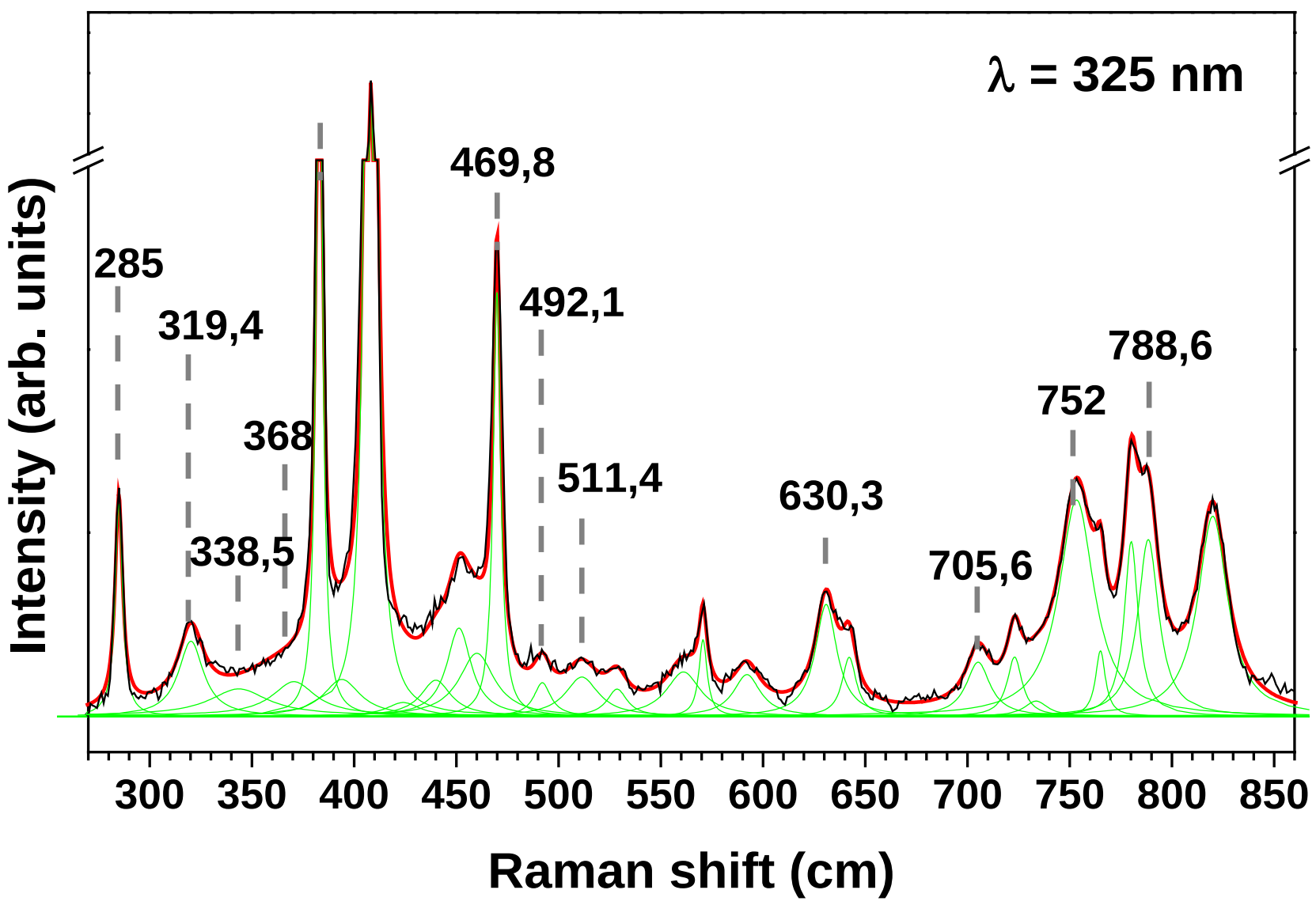

Figure S1. Raman spectrum of the bulk $MoS_2$ measured under ultraviolet (325 nm) excitation. The experimental spectrum (in dark) is fitted with Lorentzian curves (in green) and the resulting fitting curve is also shown (in red)

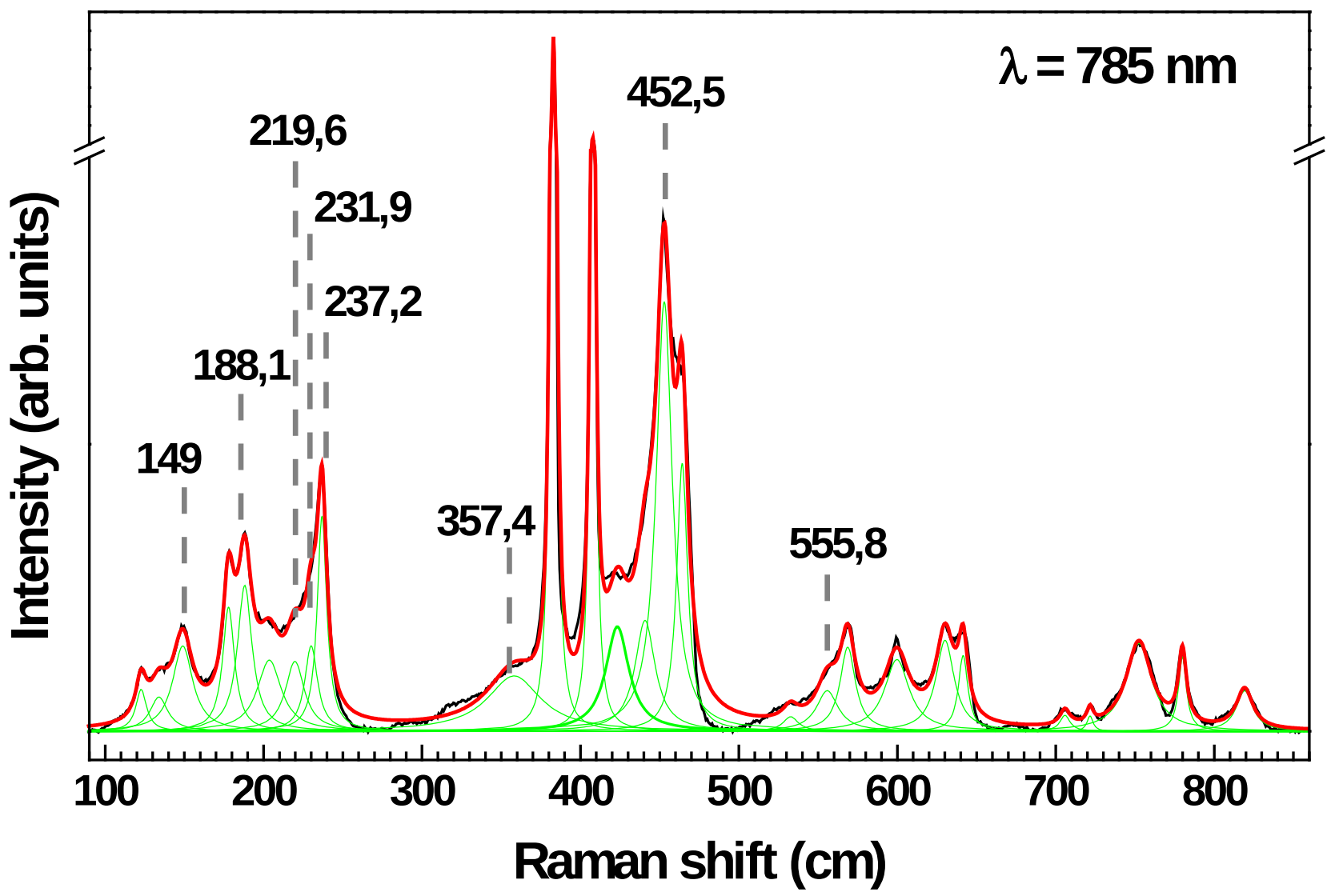

Figure S2. Raman spectrum of the bulk $MoS_2$ measured under near-infrared (785 nm) excitation. The experimental spectrum (in dark) is fitted with Lorentzian curves (in green) and the resulting fitting curve is also shown (in red)

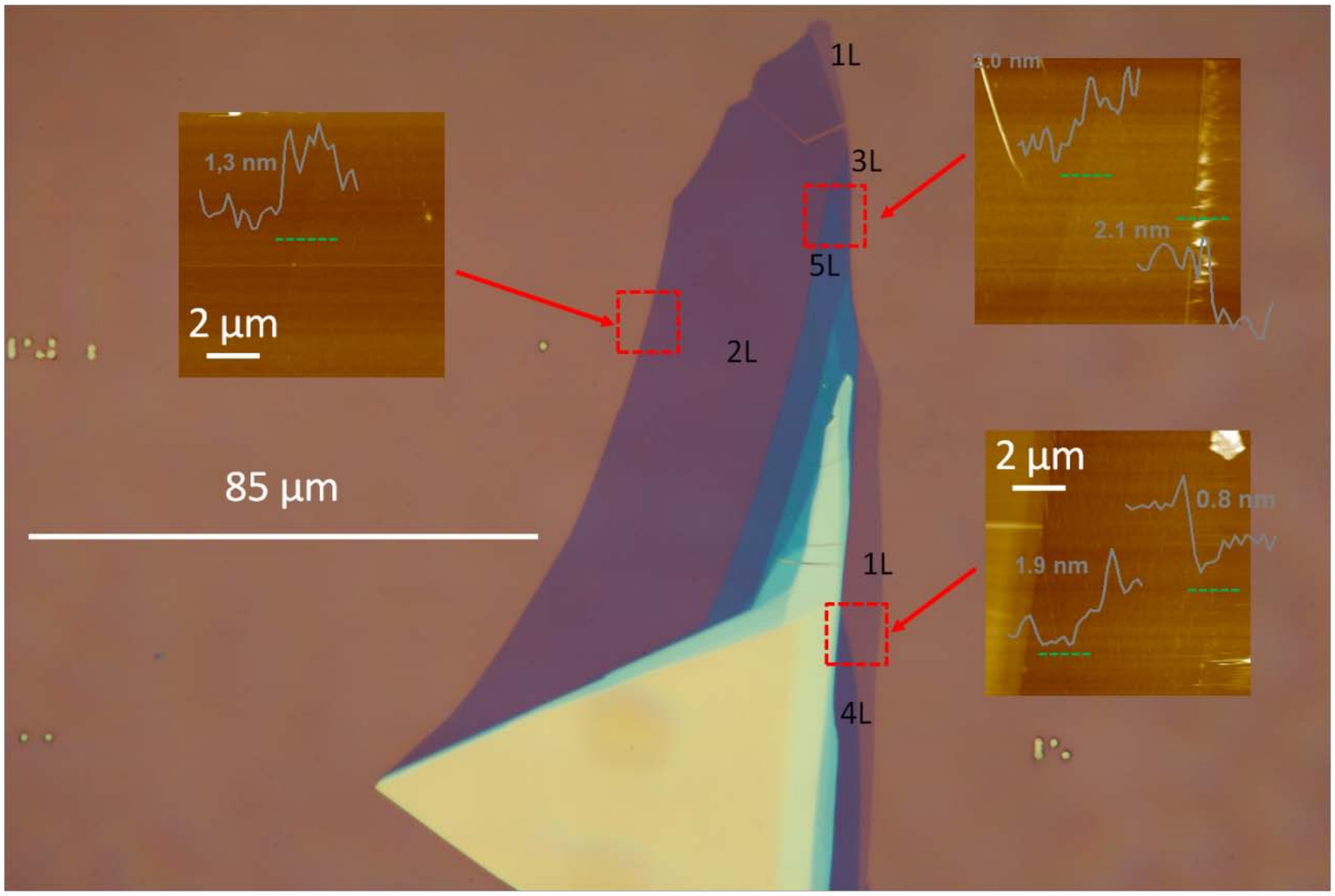

Figure S3. Optical image of the $MoS_2$ flake with the AFM scan areas. The single-layer has a thickness of ~0.8 nm, consistent with previous reports. The measured thicknesses are ~1.3 , ~2.1, ~2.7 (0.8+1.9), ~3.3 (1.3+2) nm for the 2L, 3L, 4L and 5L respectively.